\documentclass[11pt]{article}
\usepackage{hyperref}
\pdfoutput=1
\begin{document}
\title{Non-Newtonian sink flow cusps}
\author{Diego Samano \& Roberto Zenit \\
\\\vspace{6pt} Facultad de Ingenieria and Instituto de Investigaciones en Materiales \\ Universidad Nacional Aut\'onoma de M\'exico \\
Cd. Universitaria, M\'exico D.F., M\'EXICO} \maketitle
\begin{abstract}
This is an entry for the Gallery of Fluid Motion of the 62st
Annual Meeting of the APS-DFD ( fluid dynamics videos ). This
video shows the formation of non-axisymmetryc cusps in the interface of a viscoelastic liquid with air considering a selective withdrawal device.  We found that a wide variety of shapes can be observed, similar to those appearing in the rear of air bubbles ascending in non-Newtonian liquids.
\end{abstract}

\section{Introduction}

We are interested in studying non-Newtonian phenomena in general;
in particular, we have studied the shapes of free surfaces 
in complex fluids\cite{soto2006}. 

\section{Experimental Conditions}
We use a selctive withdrawal device to produce cusps in the free surface of a non-Newtonian liquid. By varying the flow rate through the tube, a variety of shapes are observed.  The liqui used in this study is an associative polymer (HASE 1.5\%). 
\section{Videos}

Our video contributions can be found at:

\begin{itemize}
    \item \href{http://hdl.handle.net/1813/14110}{Video 1, mpeg2, full
resolution}

    \item \href{http://hdl.handle.net/1813/14110}{Video 2, mpeg1, low
resolution}

\end{itemize}

\end{document}